\documentclass[mathleft]{an}
\usepackage{graphicx}
\usepackage{times}
\usepackage[varg]{txfonts}
\usepackage{color}

\def\sun{\hbox{$\odot$}}
\begin{document}

% The following seven commands are intended for editorial usage and should be ignored by
% the author(s).
\Pagespan{789}{}% Document's page range. 
% If second parameter is left empty, the last page is computed automatically.
\Yearpublication{2006}%
\Yearsubmission{2005}%
\Month{11}%   
\Volume{999}%  
\Issue{88}% 
% \DOI{This.is/not.aDOI}% 

\title{Is the semi-regular variable RU Vulpeculae undergoing a helium-shell
  flash?\thanks{Based on observations made with the Mercator Telescope, operated
    on the island of La Palma by the Flemish Community, at the Spanish
    Observatorio del Roque de los Muchachos of the Instituto de Astrof\'{i}sica
    de Canarias.}}

\author{S. Uttenthaler\inst{1}%\fnmsep\thanks{Corresponding author:
%  \email{stefan.uttenthaler@univie.ac.at}\newline}
%Example 
%for footnote, note the usage of the \texttt{fnmsep}
%command as separator between institute number and footnote mark} 
\and
R. Greimel\inst{2}
\and
M. Templeton\inst{3}
}
\titlerunning{He-shell flash in RU~Vul}
\authorrunning{S. Uttenthaler et al.}
\institute{
University of Vienna, Department of Astrophysics, T\"urkenschanzstra\ss e 17,
A-1180 Vienna, Austria;\\ \email{stefan.uttenthaler@univie.ac.at}
\and 
IGAM, Institut f\"ur Physik, Universit\"at Graz, Universit\"atsplatz 5/II,
8010 Graz, Austria;
\email{rgreimel@gmail.com}
\and 
American Association of Variable Star Observers, 49 Bay State Road, Cambridge,
MA 02138, USA;\\ \email{matthewt@aavso.org}
}

\received{03 Jul 2015}
\accepted{09 Nov 2015}
\publonline{later}

\keywords{stars: AGB and post-AGB -- stars: late-type -- stars: evolution -- stars: oscillations -- stars: individual (RU Vul)}

\abstract{
%Context:
The semi-regular variable star RU Vulpeculae (RU~Vul) is being observed visually
since 1935. Its pulsation period and amplitude are declining since $\sim1954$. A
leading hypothesis to explain the period decrease in asymptotic giant branch
(AGB) stars such as RU~Vul is an ongoing flash of the He-burning shell, also
called a thermal pulse (TP), inside the star.
%Aim:
In this paper, we present a CCD photometric light curve of RU~Vul, derive its
fundamental parameters, and test if the TP hypothesis can describe the observed
period decline.
%Methods:
We use CCD photometry to determine the present-day pulsation period and
amplitude in three photometric bands, and high-resolution optical spectroscopy
to derive the fundamental parameters. The period evolution of RU~Vul is compared
to predictions by evolutionary models of the AGB phase.
%Results:
We find that RU~Vul is a metal-poor star with a metallicity
$[{\rm M}/{\rm H}]=-1.59\pm0.05$ and an effective surface temperature of
$T_{\rm eff}=3634\pm20$\,K. The low metallicity of RU~Vul and its kinematics
indicate that it is an old, low-mass member of the thick disc or the halo
population. The present day pulsation period determined from our photometry is
$\sim108$\,d, the semi-amplitude in the V-band is $0.39\pm0.03$\,mag. The
observed period decline is found to be well matched by an evolutionary AGB
model with stellar parameters comparable to those of RU~Vul.
%Conclusions:
We conclude that the TP hypothesis is in good agreement with the observed period
evolution of RU~Vul.}
\maketitle

\section{Introduction}
The asymptotic giant branch (AGB) phase of evolution of stars with low initial
masses ($1.0 \lesssim M/M_{\sun} \lesssim 8.0$) is the last stage in which the
energy output of the star is dominated by nuclear fusion. The main source of
energy is the H-burning shell that is quasi-periodically interrupted and
shut down by the He-burning shell, which is violently ignited during
He-burning shell flashes. These events are also called thermal pulses (TPs)
and have a strong impact on many observable parameters of the star such as its
luminosity, radius, temperature, mass-loss rate, etc. For a discussion of the
evolution of stars on the AGB, see e.g.\ Habing \& Olofsson (2003).

The outer envelope of stars on the AGB pulsates on timescales of tens to
hundreds of days, thereby leading to brightness variations that are easily
detectable. Thus, these stars belong to the semi-regular or Mira-like class of
variables. Because the luminosity, radius, and temperature are altered during a
TP, also the pulsation period is expected to change considerably when the star
undergoes a TP. Wood (1975) has first suggested that these period changes may
result from the luminosity pulse produced by a TP. Wood \& Zarro (1981)
substantiated this suggestion by comparing their AGB evolutionary models to the
observed period evolution of a few Miras. Indeed, a number of Mira and
semi-regular variable (SRV) stars are known to undergo significant period
changes on timescales of decades to centuries (Zijlstra, Bedding \& Mattei 2002;
Templeton, Mattei \& Willson 2005).

One SRV with known declining pulsation period is RU~Vulpeculae (\object{RU~Vul};
Zijlstra \& Bedding 2002). RU~Vul's long-term light evolution from 1935, when
visual observations began, to 2007 was analysed by Templeton, Willson \& Foster
(2008). They find that the pulsation period started to decline around 1954,
accompanied by a decrease in amplitude and an increase in average visual
magnitude. The behaviour of RU~Vul was compared by Templeton et al.~(2008) to
that of Mira-type variables that show similarly strong period changes. In fact,
the early light curve of RU~Vul shows a quite regular variability similar to
Mira variables, only the amplitude was slightly too small to assign it to that
class. Recently, Uttenthaler et al.~(2011) obtained a high-resolution optical
spectrum of RU~Vul to search for lines of the element technetium (Tc), an
indicator of the deep mixing event called a third dredge-up (TDU) that is
expected to follow a few centuries after a TP (Vassiliadis \& Wood, 1993). No
traces of Tc were found in the spectrum of RU~Vul. However, this does not
strictly exclude the possibility of a TP going on in that star.

In recent years, the pulsation amplitude of RU~Vul has dropped to a level where
it is hardly detectable by visual estimates. We therefore decided to perform CCD
photometry to derive a reliable present-day pulsation period and amplitude in
several photometric bands. In this paper, we present the results of these
observations and determine some fundamental parameters of RU~Vul. Most
importantly, we test the prediction of the period evolution made by AGB
evolutionary models with the help of the almost 80 years of visual light curve
available for RU~Vul.

The star has been assigned to the extended (thick) disc population of the Milky
Way galaxy by Mennessier et al.~(2001), based on its kinematics. 
%This would mean the the star is fairly old ($\approx10$\,Gyrs).
We use the available information and our results to further check the population
membership of RU~Vul.

The paper is structured in the following way: In Sect.~\ref{obs} we present
the observations that were used for this study; in Sect.~\ref{analysis} the
data are analysed and results are presented; finally, in Sect.~\ref{discussion}
the results are discussed before conclusions are drawn in Sect.~\ref{conclusio}.

\section{Observations and data reduction}\label{obs}

RU~Vul is being monitored visually since 1935, and a complete visual light
curve of the star is available from the AAVSO database (Henden, 2013). We
downloaded all available data from 18 January 1935 to 15 July 2013. These data
were averaged to ten-day bins, which were then analysed with the weighted
wavelet Z-transform (wwz) algorithm developed by Foster (1996). We followed the
procedure outlined in Templeton et al.~(2005) to analyse the time-frequency
behaviour of RU~Vul. The strongest peak between 90 and 170 days of period was
adopted as the actual period.

Because in recent years the variability of RU~Vul became hardly detectable by
visual estimates, we decided to augment the available AAVSO data by CCD
photometric observations. The observations were carried out using the 30\,cm
Zeiss BMK-75 refractive telescope at Lustb\"uhel observatory operated by the
University of Graz, Austria. The data were collected using a SBIG ST-11000M CCD
camera equipped with an H$\alpha$ interference filter and Johnson-Cousins V, R
and I glass filters between 26 September 2011 (JD\,2455831.2) and 12 June 2013
(JD\,2456455.5). Standard data reduction was done using IRAF\footnote{IRAF is
distributed by the National Optical Astronomy Observatories, which are operated
by the Association of Universities for Research in Astronomy, Inc., under
cooperative agreement with the National Science Foundation.} and aperture
photometry extracted using SExtractor. The nearby K0 star HD~347112 was used as
a comparison star to convert the relative brightness measurements to actual
magnitudes. The individual measurements will be made available on the AAVSO web
page.
%Mid of our observations: JD\,2456143.35 = 03 August 2012

Furthermore, we used the ASAS V-band photometry retrieved from the ASAS project
website\footnote{{\tt http://www.astrouw.edu.pl/asas/}}. The ASAS observations
were obtained in the V band between 26 April 2003 (JD\,2452755.9) and 03
November 2009 (JD\,2455138.5).
%Start of ASAS lc: JD\,2452755.9 = 26 Apr 2003
%Mid of ASAS lc:   JD\,2453947.2 = 30 Jul 2006
%End of ASAS lc:   JD\,2455138.5 = 03 Nov 2009
%Period04 finds a best-fit period of 107.55d

The high-resolution optical spectrum of RU~Vul that we use here to derive the
effective surface temperature and metallicity was obtained on 4 July 2009 with
the fibre-fed spectrograph Hermes mounted to the 1.2\,m Mercator telescope
operated on the island of La Palma (Spain) by the Institute for Astronomy of
the University of Leuven, Belgium. The spectrograph is described in Raskin et
al.~(2010). The spectrum was initially obtained by Uttenthaler et al.~(2011),
for details we refer to that work. The spectrum covers the whole optical range
from $\sim377$ to $\sim900$\,nm at a resolving power of
$R=\lambda/\Delta\lambda=85\,000$. The typical signal-to-noise ratio is
$\sim200$ in the range used here ($\sim700$\,nm). The spectrum was
continuum-normalised in the analysed range by dividing it by a spline function
that was fit through a number of continuum points.

To derive the present-day spectral type, we also observed RU~Vul on 15 June 2012
(JD\,2456094) with the 80\,cm Cassegrain telescope of the Vienna University
Observatory using a DSS-7 spectrograph manufactured by SBIG. The visual phase
was midway from maximum to minimum at the time of observations. DSS-7 is a
grating spectrograph offering five slits with different widths, the narrowest of
which provides for a dispersion of $\sim0.55$\,nm/pixel, corresponding to a
spectral resolving power of $R\approx500$. An ST-7 CCD camera, also made by
SBIG, was connected to the spectrograph to acquire the spectra. With this
camera, the wavelength coverage is approximately $450-850$\,nm. The 2D raw
spectra were dark-subtracted and flat-fielded with dome flats. To extract the 1D
spectra, a Gaussian was fit to the profile in spatial direction. The wavelength
calibration was done using the sky emission lines by adopting a second-order
polynomial to map pixel positions to wavelength values.

\section{Analysis and results}\label{analysis}

\subsection{Photometric analysis}

\begin{figure}
\includegraphics[width=\columnwidth,bb=39 186 554 582,clip]{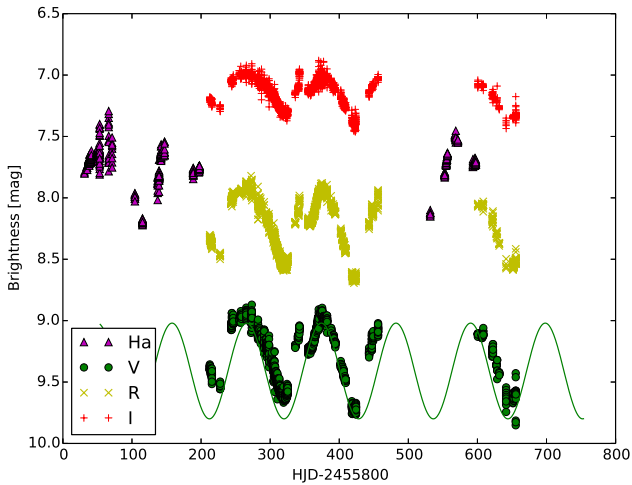}
\caption{Light curve of RU~Vul obtained at Lustb\"uhel observatory in the
  H$\alpha$ (purple triangles), V (green dots), R (yellow crosses), and I (red
  plusses) bands. The best-fit sine curve to the V-band data with a period of
  108.0\,d is also shown.}
\label{lightcurve}
\end{figure}

Our own photometric observations as well as the ASAS photometric data were
analysed with the help of the {\tt SciPy} package in Python. A least-squares fit
was applied to match a sine curve to the data. The phase, period,
semi-amplitude, and zero point offset were left free to vary.
Fig.~\ref{lightcurve} shows the light curve observed at Lustb\"uhel observatory
in the H$\alpha$, V, R, and I bands. Also shown in the figure is the best-fit
sine curve to the V-band data with a period of $108.0\pm0.5$\,d. The light curve
somewhat deviates from the sine shape and it is not strictly periodic, as
already found by Templeton et al.~(2008, their Fig.~1), which underlines
the semi-regular behaviour of RU~Vul. The phase shifts between the V, R, and I
bands were found to be insignificantly different from zero.

%%%%%%%%%%%%%%%%%%%%%%%%%%%%%%%%%%%%%%%%%%%%%%%%%%%%%%%%%%%%%%%%%%%%%%%%%%%%%%%%
%%%%%%%%%%%%%%%%%%%%%%%%%%%%%%%%%%%%%%%%%%%%%%%%%%%%%%%%%%%%%%%%%%%%%%%%%%%%%%%%

\begin{figure}
\includegraphics[width=\columnwidth,bb=68 369 535 700,clip]{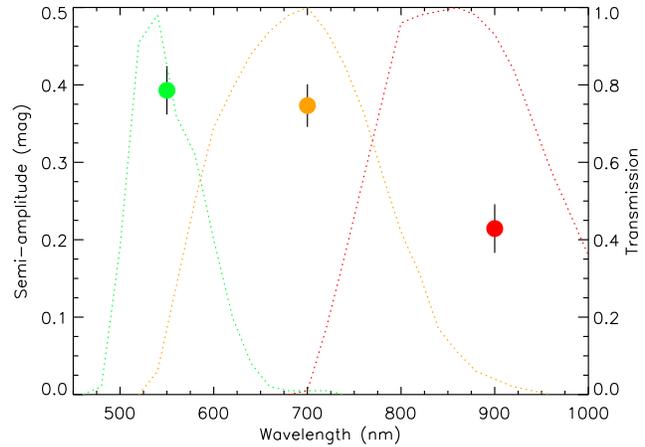}
\caption{Semi-amplitude of variation of RU~Vul as a function of central
  wavelength of the V (green), R (orange), and I band (red) filters.
%  The black symbol is
%  the amplitude found from the visual AAVSO data with the wwz algorithm.
%  The error bars in the x-axis (wavelength) subtends the range where the filter
%  transmission is $>50\%$.
  The respective filter transmission curves are plotted as dotted lines in the
  same colours, the values are to be read off on the right-hand y-axis.}
\label{ampli}
\end{figure}

The semi-amplitude of variability shows a clear dependency on the central
wavelength of the used filter. Figure~\ref{ampli} shows the measured
semi-amplitude as a function of central wavelength of the V, R, and I band
filters. The filter transmission
curves\footnote{{\tt http://ulisse.pd.astro.it/Astro/ADPS/}} are plotted as
dotted lines in the same colours as the measured amplitudes, their values are
to be read off on the right-hand y-axis. The H$\alpha$ filter is not shown here
because the amplitudes and their associated errors are derived directly from the
observed data points, but the H$\alpha$ light curve is not well sampled. The
amplitude of variation in the I band is only about half as large as in the V and
R bands.

%%%%%%%%%%%%%%%%%%%%%%%%%%%%%%%%%%%%%%%%%%%%%%%%%%%%%%%%%%%%%%%%%%%%%%%%%%%%%%%%
%%%%%%%%%%%%%%%%%%%%%%%%%%%%%%%%%%%%%%%%%%%%%%%%%%%%%%%%%%%%%%%%%%%%%%%%%%%%%%%%

%Interestingly, the V band amplitude found by our photometric observations
%($0.786\pm0.062$\,mag) is much larger than the one found from the concurrent
%visual AAVSO data using the wwz algorithm ($\sim0.17$\,mag). The amplitude
%found in the photometric measurements should actually be easily detectable by
%visual estimates (Lebzelter \& Kiss, 2001). The reason for this discrepancy
%is unclear, but it seems that too many visual observers spoil the amplitude in
%this case. Inspecting the estimates from different observers indicates that
%they do have considerable offsets.
%
%We note that the visual amplitude obtained via wavelet analysis is less than
%the V amplitude; even accounting for the fact that WWZ reports the
%semi-amplitude rather than the peak-to-trough range, the visual range is about
%half that of the V-band amplitude, and closer to (but still less than) the
%I-band amplitude. There may be multiple effects occurring with visual data,
%including observers using different sequences, having different visual
%acuities, and using different observing methods and equipment; the visual
%bandpass is also wider and has a somewhat redder bias than Johnson V, which may
%also account for some of the difference. In any case, the visual data match the
%instrumental data in overall behavior, but the amplitudes observed do not.

Interestingly, the V band semi-amplitude found by our photometric observations
indicates that the variability of RU~Vul should be large enough to be easily
detectable by visual eye estimates (Lebzelter \& Kiss, 2001), but it is hardly
revealed. The range of reported brightnesses is in good agreement with that
found by the photometric observations, but there is no regular variability in
the light curve of the visual observations. Only if specific individual
observers are picked out, a pulsation pattern is revealed. The reason for this
is unclear to us.

\begin{figure}
\includegraphics[width=\columnwidth,bb=38 188 552 582,clip]{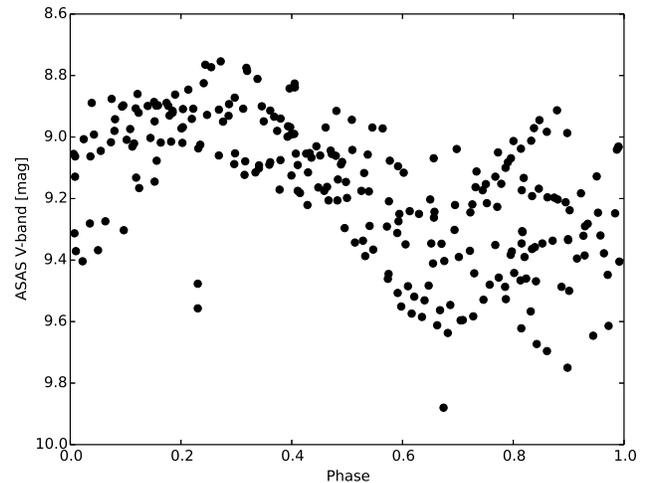}
\caption{Folded light curve of the ASAS photometric data, using a period of
  107.6\,d.}
\label{ASAS}
\end{figure}

Finally, also the ASAS V band light curve was analysed in the same way as our
own photometric observations. A best-fit period of 107.6\,d was found, a
somewhat lower value than the 113.1\,d quoted in the ASAS catalogue itself.
However, inspection of the (folded) light curve shows that the value found by us
appears to be a much better match to the observations, see Fig.~\ref{ASAS}. The
pulsation period derived from ASAS is identical with that from our own
observations within the uncertainties, indicating that the period has not
significantly declined within the approximately six years between the ASAS and
our observations.% Figure~\ref{ASAS} shows the ASAS light curve folded with the
%period of 107.6\,d derived here.

%%%%%%%%%%%%%%%%%%%%%%%%%%%%%%%%%%%%%%%%%%%%%%%%%%%%%%%%%%%%%%%%%%%%%%%%%%%%%%%%
%%%%%%%%%%%%%%%%%%%%%%%%%%%%%%%%%%%%%%%%%%%%%%%%%%%%%%%%%%%%%%%%%%%%%%%%%%%%%%%%

\subsection{Spectroscopic analysis}

To derive the effective surface temperature and the metallicity of RU~Vul from
the high-resolution optical spectrum, we applied a similar approach as Valenti,
Piskunov \& Johns-Krull (1998). These authors used an interpolation and fitting
routine in a three-dimensional grid of synthetic spectra in
$(T_{\rm eff},[{\rm M}/{\rm H}],\log g)$ to derive the fundamental parameters
of an M-type dwarf star. Here, we use an independent estimate of the surface
gravity $\log g$ and do the interpolation and fitting only in a
two-dimensional grid $(T_{\rm eff},[{\rm M}/{\rm H}])$.

To obtain an estimate of the surface gravity, we adopt the luminosity of
$L=2830L_{\sun}\pm520L_{\sun}$ given by Uttenthaler et al.~(2011). This was
derived from a period-magnitude relation for the K-band (sequence~1 for O-rich
stars in Riebel et al.~2010, their Table~6) and a bolometric correction to the
absolute K magnitude based on the $(J-K)_0$ colour (Kerschbaum, Lebzelter
\& Mekul 2010, their Group~A). The uncertainty on the luminosity was determined
by combining in quadrature the uncertainties stemming from the 2MASS photometry
(which is however minor) and the uncertainties quoted for the coefficients of
the period-magnitude relation and the BC(K) correction. Since RU~Vul probably
is a member of the old, thick disc component of the Galaxy (Mennessier et
al.~2001), we adopt a mass of $1M_{\sun}$ for RU~Vul. Considerations of the
stellar life-time show that this is probably a good approximation. Furthermore,
Uttenthaler et al.~(2011) estimated a temperature of 3700\,K from the mean
spectral type (M3). Putting together the ingredients for stellar mass,
luminosity, and temperature, we calculate $\log g = 0.22$.

Given the surface gravity, we established a grid of model atmospheres with the
COMARCS code (Aringer et al.~2009) and synthetic spectra based on this
atmosphere grid. The grid covered the range 3500 -- 4100\,K in steps of 100\,K
and $[{\rm M}/{\rm H}]=-2.0$ to $-0.5$ in steps of 0.25\,dex. Uttenthaler et
al.~(2011) had estimated the temperature at 3700\,K and the metallicity to
be close to $[{\rm M}/{\rm H}]=-1.5$, so the grid was constructed to bracket
these approximate values. Only scaled solar abundances were used, assuming the
set of solar abundances of Caffau et al.~(2008). Excessive atomic (VALD; Kupka
et al.~2000) and molecular (TiO, Schwenke 1998) line lists were used in the
synthesis of the spectra with the COMA code (Aringer 2000). We used the range
698 -- 715\,nm for the fitting. This wavelength range is constrained by telluric
lines that contaminate the observed spectrum at shorter and longer wavelengths.
This piece of spectrum contains the TiO $\gamma(0,0)Ra$, $\gamma(0,0)Rb$, and
$\gamma(0,0)Rc$ band heads, which are sensitive to the surface temperature of
cool red giants. The metallicity is constrained simultaneously by the atomic
lines that are present in that spectral range; in RU~Vul, the TiO bands are not
too strong to blanket all atomic lines, thus only one piece of spectrum is
required to derive both quantities\footnote{Valenti et al. (1998) used two
separate pieces of spectrum, one dominated by the temperature-sensitive TiO
bands, another one dominated by metallicity-sensitive atomic lines.}.

The spectra were synthesised with an original resolution of $R=3\times10^5$,
assuming a micro-turbulent velocity of 2.5\,km\,s$^{-1}$. The synthetic
spectra were smoothed to the resolution of 85\,000 of the observed spectrum
and different values of macro-turbulent velocities were added to improve the
fit. It turned out that a macro-turbulent velocity of 11\,km\,s$^{-1}$
provided the best fit, but that the precise choice had virtually no influence
on the obtained best-fit combination of temperature and metallicity; it only
had the effect of improving the quality of the fit. Following Valenti et
al.~(1998), we did the interpolation in $\log(1-f)$, where $f$ is the
continuum-normalised flux, to improve the convergence of the fitting routine.
A custom-made IDL routine was programmed, using the {\tt amoeba.pro} simplex
routine to find the minimum of a $\chi^2$ figure of merit. Because of inaccurate
data in the atomic line list, in some lines the synthetic flux deviated
considerably from the observed one, so we omitted wavelength points from the
fitting procedure if $f_{\rm obs}-f_{\rm synth}$ was larger than $0.1f_{\rm obs}$,
where $f_{\rm obs}$ is the observed flux and $f_{\rm synth}$ the synthetic model
flux. We found that the fitting routine converged reliably to the same best-fit
combination of temperature and metallicity independent from the starting values
within the grid.

\begin{figure}
  \includegraphics[width=\columnwidth,bb=84 370 548 700,clip]{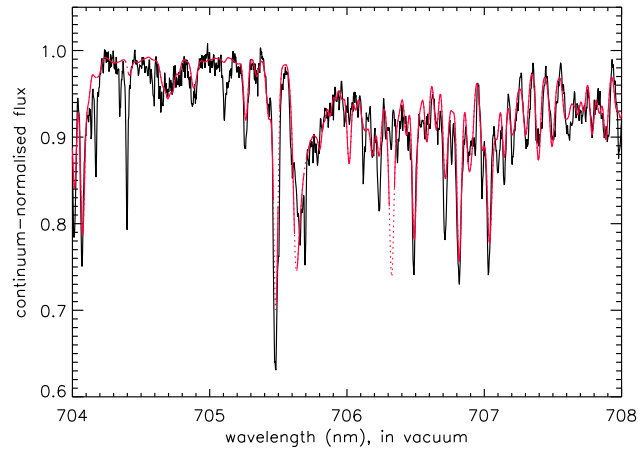}
  \caption{Portion of the spectral range that was used to determine the
    metallicity and temperature of RU~Vul. The observed spectrum is plotted as
    black line, the best-fitting interpolated synthetic spectrum as red
    line. Data points that were not included in the fitting procedure because
    they deviate by more than 10\% from the observed flux are shown as dotted
    line. The broad feature starting at 705.6\,nm is the TiO $\gamma(0,0)Ra$
    band head. Note the zoomed range on the y-axis.}
  \label{Teff_MH_best-fit}
\end{figure}

Eventually, we found that the best-fit combination of temperature and
metallicity or RU~Vul is $T_{\rm eff}=3634\pm20$\,K and
$[{\rm M}/{\rm H}]=-1.59\pm0.05$. An illustration of the observed spectrum
together with the best-fit interpolated synthetic spectrum is shown in
Fig.~\ref{Teff_MH_best-fit}. Overall, the fit is very satisfying and the
determination of temperature and metallicity seem to be reliable.

With the newly determined temperature, the surface gravity changes to
$\log g = 0.18$ and one could do another iteration step to refine the
metallicity and temperature by establishing a new model grid. However, since
the change in $\log g$ is small and the features in the investigated spectral
range are quite insensitive to $\log g$, we omitted another iteration.

%%%%%%%%%%%%%%%%%%%%%%%%%%%%%%%%%%%%%%%%%%%%%%%%%%%%%%%%%%%%%%%%%%%%%%%%%%%%%%%%
%%%%%%%%%%%%%%%%%%%%%%%%%%%%%%%%%%%%%%%%%%%%%%%%%%%%%%%%%%%%%%%%%%%%%%%%%%%%%%%%

Table~\ref{params} summarizes the most important stellar parameters derived
here.

%%%%%%%%%%%%%%%%%%%%%%%%%%%%%%%%%%%%%%%%%%%%%%%%%%%%%%%%%%%%%%%%%%%%%%%%%%%%%%%%
\begin{table}
% \centering%%%
\caption{Main parameters of RU~Vul.}
\label{params}
\begin{tabular}{cc}
\hline
Parameter & Value \\
%(unit1) & (unit2) \\
\hline
distance (pc)                 & $2070\pm130$ \\
luminosity ($L_{\sun}$)        & $2830\pm520$ \\
$\log (g \rm{[cm s^{-2}]})$    & 0.18 \\
$T_{\rm eff}$ (K)               & $3634\pm20$ \\
$[{\rm M}/{\rm H}]$           & $-1.59\pm0.05$ \\
radial velocity (km\,s$^{-1}$) & $-67.9$ \\
pulsation period (d)          & $\sim108.0$ \\
$<V>$ (mag)                   & 9.329 \\
$V$ amp.\ (mag)               & $0.393\pm0.031$ \\
$<R>$ (mag)                   & 8.287 \\
$R$ amp.\ (mag)               & $0.373\pm0.028$ \\
$<I>$ (mag)                   & 7.196 \\
$I$ amp.\ (mag)               & $0.214\pm0.031$ \\
\hline
\end{tabular}
\end{table}
%%%%%%%%%%%%%%%%%%%%%%%%%%%%%%%%%%%%%%%%%%%%%%%%%%%%%%%%%%%%%%%%%%%%%%%%%%%%%%%%

\section{Discussion}\label{discussion}

\subsection{Population membership}

Mennessier et al.\ (2001) assigned RU~Vul to the extended (thick) disc. With a
metallicity of $[{\rm M}/{\rm H}]=-1.59$, RU~Vul is more metal-poor than most of
the thick disc stars. However, it was shown by Ruchti et al.~(2010) and Ruchti
et al.~(2011) that thick disc stars can have $[{\rm Fe}/{\rm H}]<-2$, although
these are few objects. On the other hand, with its metallicity RU~Vul is just
at the peak metallicity of the metal-rich inner halo (Carollo et al., 2007).
Note that we measured a general metallicity [M/H], not [Fe/H]. Because
metal-poor thick disc and halo stars have a general over-abundance of
$\sim0.3$\,dex in the $\alpha$-elements, to which also O and Ti belong, [Fe/H]
may be lower than [M/H] in RU~Vul. According to Reddy \& Lambert~(2008),
metal-weak thick disc stars cannot be assigned to a particular component.

Using the radial velocity of $-67.9$\,km\,s$^{-1}$ and distance of
2070\,pc from Uttenthaler et al.~(2011) and proper motion
$({\rm pm}_{\rm RA},{\rm pm}_{\rm DE})=(1.13,6.31)$\,mas/yr from van Leeuwen
(2007), we determine a space velocity of
$(u,v,w)=(-76.5,-33.6,39.8)$\,km\,s$^{-1}$. With this, RU~Vul falls among
the thick disc stars in the Toomre diagram of Bensby et al.~(2005).

Its proximity to the Galactic plane ($b=-10\degr916$, $z=-518$\,pc) speaks in
favour of a disc membership, but it could also be a halo star that is close to
the plane by chance. Thus, a clear population membership cannot be assigned to
RU~Vul, both a membership in the extended (thick) disc or in the metal-rich
inner halo seem possible.
%$Z_{\rm max}$ from Marica.
%Membership in the thick disc or halo population: Check with kinematics!
%$RV=-67.9$\,km\,s$^{-1}$, pm_RA=1.13 mas/yr, pm_DE=6.31 mas/yr (van Leeuwen
%2007), this varies a lot!
%RA=20 38 52.68850, Dec=+23 15 31.1862
%d=2684 pc ; Wrong distante !!!!!!!!!!!!!!!!!!!!!!!!!!!!!!!!!!!!!!!!!!
%GAL_UVW, U, V, W, RA=309.71954d0, DEC=23.258663d0, PMRA=1.13d0 ,PMDEC=6.31d0, VRAD=-67.9d0, DISTANCE=2684.0d0
%u=91.256546, v=-25.452619, w=47.824430 km/s
%GAL_UVW, U, V, W, RA=309.71954d0, DEC=23.258663d0, PMRA=1.13d0 ,PMDEC=6.31d0, VRAD=-67.9d0, DISTANCE=2684.0d0,/LSR
%u_lsr=82.756546, v_lsr=-12.072619, w_lsr=54.314430 km/s
%%%%%%%%%%%%%%%%%%%%%%%%%%%%%%%%%%%%%%%%%%%%%%%%%%%%%%%%%%%%%%%%%%%%%%%%%%%%%%%%
%d=2070.0d0 pc
%GAL_UVW, U, V, W, RA=309.71954d0, DEC=23.258663d0, PMRA=1.13d0 ,PMDEC=6.31d0, VRAD=-67.9d0, DISTANCE=2070.0d0
%u=76.499767, v=-33.600531, w=39.825446

\subsection{Comparison to AGB evolutionary models}

\begin{figure}
  \includegraphics[width=\columnwidth,bb=78 370 538 700,clip]{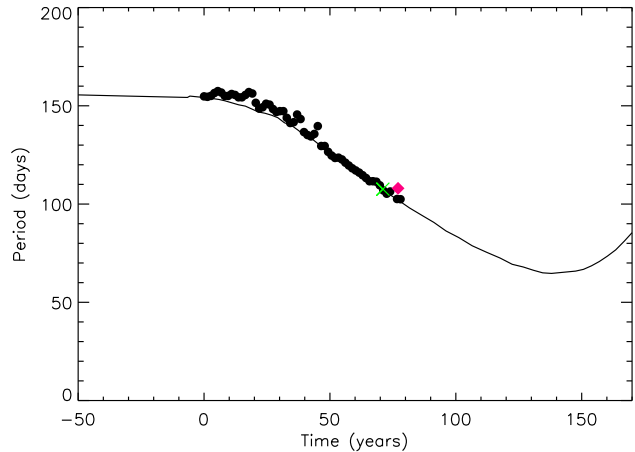}
  \caption{Period evolution of RU~Vul between 1935 and 2013 (black dots) and of
    the chosen AGB evolutionary model from Vassiliadis \& Wood (1993;
    $1M_{\sun}$, $Z=Z_{\sun}/16$, black line) at the beginning of the sixth TP of
    that model. Also shown is the period determination from the ASAS
    observations (green cross) as well as our own observations (red diamond).}
  \label{P_evol}
\end{figure}

Next we confront AGB evolutionary models with the observed period evolution of
RU~Vul. We selected a model from Vassiliadis \& Wood (1993) that well represents
the main parameters of that star, namely the model with a mass of $1M_{\sun}$ and
a metallicity of $Z=Z_{\sun}/16$, corresponding to $[{\rm M}/{\rm H}]=-1.20$,
which is the model with the lowest metallicity in that set. The pulsation period
in days was calculated from the total mass and the stellar model radius (derived
from luminosity and surface temperature) using the equations given in
Vassiliadis \& Wood (1993). From that model, we selected a thermal pulse cycle
for comparison with the observations where the pulsation period before the onset
of the TP (i.e.\ at the end of the quiescent H-burning phase) is close to the
period of RU~Vul before the period decrease, namely $\sim155$\,d. This is the
case for the sixth TP (out of eleven TPs) of this model. We applied an arbitrary
shift on the zero point of the time axis 
%of 3\,885\,068 years
to match the observed period decline with that of the model. The result of this
exercise is presented in Fig.~\ref{P_evol}, which also includes the period
determination from the ASAS and our own data. The period evolution of RU~Vul
until 2007 has already been presented by Templeton et al.\ (2008, see their
Fig.~2), here we only present an update of those data until 2013. The match
between observed and predicted period evolution is remarkably good. We emphasize
that except selecting this particular model and particular TP and the time zero
point, we did not apply any fitting of the model to the data. However, it needs
to be noted that a broad range of observed period behaviour may be fitted by
different evolutionary models, thus a good match between observed and modelled
period evolution is no solid proof that indeed a TP is going on in a given star.

%%%%%%%%%%%%%%%%%%%%%%%%%%%%%%%%%%%%%%%%%%%%%%%%%%%%%%%%%%%%%%%%%%%%%%%%%%%%%%%%
%%%%%%%%%%%%%%%%%%%%%%%%%%%%%%%%%%%%%%%%%%%%%%%%%%%%%%%%%%%%%%%%%%%%%%%%%%%%%%%%

Nevertheless, the model uncertainties on the short-time evolution are expected
to be small because the models are quite robust regarding structural changes
during a TP. Uncertainties in the treatment of mass loss and convection are not
expected to alter the results significantly on a relatively short time scale of
several decades.

%\smallskip
%The observed period evolution of RU~Vul is in excellent agreement with the
%evolution predicted by an AGB model with similar parameters. 
%Thus, assuming
%that RU~Vul is undergoing a TP, the observations confirm the AGB evolutionary
%models of Vassiliadis \& Wood (1993) used here. Vice versa, if we assume that
%the AGB models give an accurate prediction of the period evolution during a TP,
%we may conclude that RU~Vul is undergoing a TP. Since we do not have any
%independent means to determine if the star is indeed undergoing a TP, we cannot
%draw any conclusion stronger than this.
%
%Peter: I think the models are quite robust regarding structural changes
%during a thermal pulse. I doubt if mass loss or convection treatment will
%significantly alter the results.

\begin{figure}
  \includegraphics[width=\columnwidth,bb=70 368 502 850,clip]{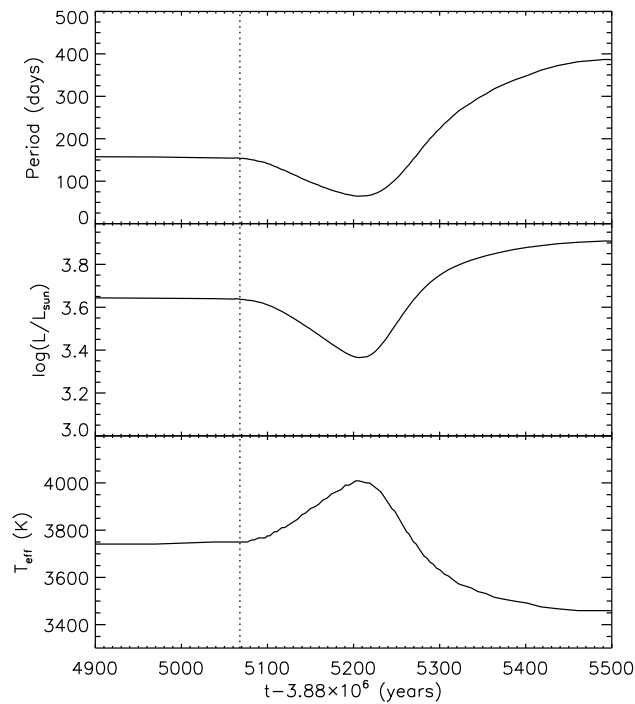}
  \caption{Evolution of pulsation period (upper panel), luminosity (middle
    panel), and effective temperature (lower panel) of the model used in
    Fig.~\ref{P_evol} during the first few hundred years after the onset of the
    6th TP. The time is measured since the beginning of the AGB. The vertical
    dotted line marks the time zero point chosen for Fig.~\ref{P_evol}.}
  \label{mod_ev}
\end{figure}

Besides the pulsation period, which is a measure for the stellar radius, the TP
scenario can be compared to other stellar quantities for further constraints.
Figure~\ref{mod_ev} shows the evolution of the pulsation period, effective
luminosity, and temperature of the adopted model for the first few hundred years
after the onset of this particular TP. The model predicts that the pulsation
period and the luminosity start to decrease when a TP begins, reaching minima of
$\sim65$\,d and $\sim2300L_{\sun}$ some 135 years after the onset of the TP,
whereas the temperature starts to rise, reaching a maximum of $\sim4000$\,K.
Later in the evolution of the TP, these trends reverse (Fig.~\ref{mod_ev}).
We inspect the available observational material of RU~Vul to put further
constraints on the TP hypothesis.

%%%%%%%%%%%%%%%%%%%%%%%%%%%%%%%%%%%%%%%%%%%%%%%%%%%%%%%%%%%%%%%%%%%%%%%%%%%%%%%%
%%%%%%%%%%%%%%%%%%%%%%%%%%%%%%%%%%%%%%%%%%%%%%%%%%%%%%%%%%%%%%%%%%%%%%%%%%%%%%%%

The luminosity of the model star at the end of quiet H-burning is
$\sim4350L_{\sun}$, compared to $\sim2830\pm520L_{\sun}$ determined for RU~Vul.
This constitutes maybe the most serious difference between the actual properties
of RU~Vul and the adopted model, but note that the observed luminosity is based
on a period-luminosity relation (using the period before the decline), not on an
actually measured distance. Nevertheless, since the period decrease has
commenced, RU~Vul has {\it brightened} in the visual range because the cycle
minima have become brighter, while maxima have roughly remained constant
(Templeton et al.~2008, their Fig.~1). However, the visual (or V band)
magnitude is not a good indicator for the bolometric luminosity of a red giant
star because most of its light is emitted in the near IR. Unfortunately, no
early IR observations of RU~Vul are available.

The surface temperature of the model has risen from $\sim3750$\,K to
$\sim3870$\,K by 78 years after the onset of the TP. This rise in temperature
could lead to the observed increase in the visual brightness of RU~Vul due to
two effects. First, the emission maximum will be shifted to shorter wavelengths
following the Wien law. Secondly, a higher temperature will decrease the
molecular opacity in the visual range, which is mainly due to TiO. %This second
%effect might not be very strong in RU~Vul due to its low metallicity.
The increase in temperature may also explain the observed decrease in pulsation
amplitude (Templeton et al.~2008, their Fig.~2) because at a higher temperature
the molecular TiO bands will be weaker.

An indication of an increase in temperature might come from spectral type
classifications listed by Skiff (2013). RU~Vul has been classified as M3e in
1897, M2 prior to 1943, and M4e in 1958, just as the period decline has started.
We compared our own low-resolution observations (Sect.~\ref{obs}) to the
spectral type standards of Fluks et al.~(1994) and find that the current
spectral type is M0. This may indicate that the surface temperature of RU~Vul
has somewhat increased in recent decades, in agreement with expectations. We
thus think that the brightness increase in the visual range is not in
contradiction with the TP hypothesis.

As reported by Uttenthaler et al.~(2011), no Tc has been detected in the
atmosphere of RU~Vul. Technetium is expected to be found on the surface of AGB
stars that underwent a TDU, a deep mixing event that is thought to occur after
powerful TPs. The simplest explanation for the absence of Tc is that no TDU
event has (yet) taken place in RU~Vul, or that the star has too low a mass to
ever undergo TDU. On the other hand, very few seed nuclei are available at
low metallicity so that the light s-process elements such as Tc are only very
little enhanced, in favour of the heavy s-elements (Busso, Gallino \&
Wasserburg 1999). However, also the singly-ionized lines of Ba at 649.87\,nm and
of La at 652.89\,nm indicate that these s-elements are not enhanced in strength.
Another effect of TDU would be the enrichment of the atmosphere with carbon,
which would quickly lead to a C/O ratio by number in excess of unity at low
metallicity. Since RU~Vul is clearly oxygen-rich, as signalled by the presence
of TiO bands, we conclude that TDU has not occurred in RU~Vul.
% La and Ba seem to be not enhanced!

Finally, there are also other mechanisms that have been proposed to be able to
lead to relatively quick changes of pulsation period in red giant stars.
Zijlstra et al.~(2004) propose that a possibly chaotic feed-back between
molecular opacities, pulsation amplitude, and period can cause an unstable
period in stars with a C/O ratio very close to unity (SC and CS spectral types),
as small changes in the temperature in the atmosphere can cause large changes in
the molecular abundances, and hence the opacities. This mechanism can be
excluded to be at play in RU~Vul because the C/O ratio is clearly below unity.
Furthermore, Ya’ari \& Tuchman (1996) and Lebzelter \& Wood (2005) identify a
feedback mechanism between the pulsation and the stellar entropy structure,
which can lead to mode switching. However, there are no detailed predictions as
to how these mechanisms could be identified observationally.

\section{Conclusions}\label{conclusio}

We find that the SRV RU~Vul is a metal-poor ($[{\rm M}/{\rm H}]=-1.59\pm0.05$),
cool ($T_{\rm eff}=3634\pm20$\,K) red giant star whose chemical and kinematic
properties are in agreement with either a thick-disc or a halo membership. The
pulsation period decrease observed in RU~Vul is well described by evolutionary
models of the TP-AGB matched to the properties of the star. A decrease of the
pulsation period for the next few decades is expected from these models.
Also other observables such as pulsation amplitude and spectral sub-type, a
proxy for the surface temperature, are in agreement with expectations from an
ongoing TP. This suggests that RU~Vul is a good candidate for a star that
currently undergoes a TP, though our work cannot provide a solid proof for this
suggestion. The evolution of this star is worthwhile to follow to learn more
about these short, critical phases close to the endpoint of low-mass stellar
evolution. Finally, we propose that RU~Vul would also be of interest to study
the mass-loss mechanism in metal-poor, pulsating red giants.
\vspace{1cm}

\acknowledgements
SU acknowledges support from the Austrian Science Fund (FWF) under project
P~22911-N16. We thank Stefan Meingast for the assistance in the observations
at the Vienna University Observatory. We acknowledge with thanks the variable
star observations from the AAVSO International Database contributed by observers
worldwide and used in this research.

%%%%%%%%%%%%%%%%%%%%%%%%%%%%%%%%%%%%%%%%%%%%%%%%%%%%%%%%%%%%%%%%%%%%%%%%%%%%%%%%

%%%%%%%%%%%%%%%%%%%%%%%%%%%%%%%%%%%%%%%%%%%%%%%%%%%%%%%%%%%%%%%%%%%%%%%%%%%%%%%%

\end{document}